\documentclass[journal]{IEEEtran}

\usepackage[pdftex]{graphicx}
\DeclareGraphicsExtensions{.pdf,.jpeg,.png}

\usepackage{amsmath,amsfonts}
\usepackage{amsthm}
\usepackage{amssymb}
\usepackage{algorithmic}
\usepackage{algorithm}
\usepackage{aligned-overset}
\usepackage{array}
\usepackage{balance}
\usepackage{booktabs}
\usepackage{cite}
\usepackage{color}
\usepackage{epstopdf}
\usepackage{enumitem}
\usepackage{mathtools}
\usepackage{bbm}
\usepackage{makecell}
\usepackage{multicol}
\usepackage{multirow}
\usepackage[caption=false,font=normalsize,labelfont=sf,textfont=sf]{subfig}
\usepackage{tabularx}
\usepackage{textcomp}
\usepackage{stfloats}
\usepackage{url}
\usepackage{verbatim}
\usepackage{graphicx}
\usepackage{url}

\usepackage{etoolbox}

\graphicspath{{./Figures/}}

\newtheorem{lemma}{Lemma}

\newtheorem{Theorem}{Theorem}

\renewcommand{\vec}[1]{\boldsymbol{\mathrm{#1}}}

\newcommand{\src}{\textnormal{S}}
\newcommand{\uav}{\textnormal{U}}
\newcommand{\ris}{\textnormal{R}}
\newcommand{\des}{\textnormal{D}}

\newcommand{\XY}{\textnormal{XY}}
\newcommand{\nodeX}{\textnormal{X}}
\newcommand{\nodeY}{\textnormal{Y}}

\newcommand{\ccia}{\textnormal{I}}
\newcommand{\ccig}{\textnormal{J}}
\newcommand{\sinr}{{\Gamma}}
\newcommand{\pl}{{\ell}}

\newcommand{\chain}{{\Phi}}

\newcommand{\atog}{\mathtt{a2g}}
\newcommand{\gtoa}{\mathtt{g2a}}
\newcommand{\cas}{\mathtt{cas}}
\newcommand{\dir}{\mathtt{dir}}

\newcommand{\tsqrt}[1]{{\textstyle{\sqrt{#1}}}}
\newcommand{\tsum}{\mathop{\textstyle\sum}}
\newcommand{\tprod}{\mathop{\textstyle\prod}}

\newcommand{\sprod}{\mathop{\scriptstyle\prod}}

\usepackage{titlesec}
\BeforeBeginEnvironment{align}{\vspace{-5pt}}
\titlespacing*{\subsection}{5pt}{5pt}{5pt}
\titlespacing*{\section}{3pt}{3pt}{3pt}

\begin{document}
\allowdisplaybreaks

\title{Channel Characterization of UAV-RIS-aided Systems with Adaptive Phase-shift Configuration}

\author{ Thanh Luan~Nguyen,     Georges~Kaddoum,~\IEEEmembership{Senior~Member,~IEEE,}~Tri~Nhu~Do,~and~Zygmunt~J.~Haas,~\IEEEmembership{Fellow,~IEEE}

    \thanks{This work was supported by the Tier 2 Canada Research Chair program and the Natural Sciences and Engineering Research Council of Canada (NSERC) Discovery Grant program. The work of Z. J. Haas was supported in part by the U.S. National Science Foundation under the grant number CNS-1763627.}
    \thanks{Thanh~Luan~Nguyen, Georges~Kaddoum and Tri~Nhu~Do are with the Department of Electrical Engineering, the \'{E}cole de Technologie Sup\'{e}rieure (\'{E}TS), Universit\'{e} du Qu\'{e}bec, Montr\'{e}al, QC H3C 1K3, Canada. Georges~Kaddoum is also with Cyber Security Systems and Applied AI Research Center, Lebanese American University (emails: thanh-luan.nguyen.1@ens.etsmtl.ca, georges.kaddoum@etsmtl.ca, tri-nhu.do@etsmtl.ca).}	
    \thanks{Z.~J.~Haas is with the Department of Computer Science, University of Texas at Dallas, Richardson, TX 75080, USA, and also with the School of Electrical and Computer Engineering, Cornell University, Ithaca, NY 14853, USA (e-mail: zhaas@cornell.edu).}	
}
		
\maketitle

\begin{abstract}
This letter considers a UAV aiding communication between a ground transmitter and a ground receiver in the presence of co-channel interference. A discrete-time Markov process is adopted to model the complex nature of the Air-to-Ground (A2G) channel, including the occurrence of Line-of-Sight, Non-Line-of-Sight, and blockage events. Moreover, an adaptive phase-shift-enabled Reconfigurable Intelligent Surface (RIS) is deployed to combat A2G blockage events. Novel frameworks based on the shadowed Rician distribution are proposed to derive closed-form expressions for Ground-to-Air/A2G SINR' distributions. Numerical results show that RISs with large numbers of elements, e.g., 256 RIS elements, improve end-to-end Outage Probability (OP) and reduce blockages.
\end{abstract}

\begin{IEEEkeywords}
Reconfigurable Intelligent Surfaces (RIS); UAV; adaptive phase-shift configuration; channel characterization
\end{IEEEkeywords}

\IEEEpeerreviewmaketitle

\section{Introduction} \label{section_intro}

\IEEEPARstart{R}{econfigurable} intelligent surfaces (RISs) have emerged as a promising technology for enhancing the communication performance of non-terrestrial networks \cite{LinTAES2022}, especially in the context of Unmanned Aerial Vehicles (UAVs)-based wireless communication systems~\cite{ChenCM2023}. 
    RISs, composed of programmable RIS elements, can manipulate the direction of incident signals, improve the received signal strength, mitigate interference, and enhance coverage rate.
These advantages make RISs specifically useful in UAVs-based systems to enable remote communication even in interference-limited and disaster-stricken areas.
    However, the inherently dynamic nature of UAVs presents distinct challenges when it comes to accurately characterizing both Air-to-Ground (A2G) and Ground-to-Air (G2A) communication~channels.

In \cite{SalhabWCL2021}, the authors provided an accurate Outage Probability (OP) expression for RIS-aided interference-free wireless networks under Rician fading. Moreover, \cite{BoulogeorgosTVT2022} recently presented closed-form OP expressions considering disorientation, misalignment, and hardware imperfections. Similarly, \cite{BianOJCS2023} recently derived closed-form expressions for OP and channel capacity in interference scenarios. The work in \cite{YangTVT2020} expanded this focus, investigating performance in UAV-RIS-aided interference-free networks. Alongside these, \cite{JiangTC2023} introduced a three-dimensional (3D) LoS/NLoS channel model for UAV-to-ground communications, using a dual RIS setup on building facades. Finally, \cite{SavkinCL2023} contributed a new UAV deployment algorithm for 3D non-homogeneous terrain.
    While the studies in \cite{SalhabWCL2021, BoulogeorgosTVT2022, BianOJCS2023, YangTVT2020, JiangTC2023} consider LoS and NLoS conditions, the concurrent blockages that disrupt communication is overlooked. Our work bridges this gap by deriving a comprehensive model accounting for LoS, NLoS, and blockages, aiming to accurately represent challenges in UAV-RIS-aided network performance.
    
Different from existing works, in our work, the elevation angle in A2G communication links can experience stochastic behaviors, leading to significant impacts on the channel quality. 
    Specifically, variations in the elevation angle can lead to unpredictable blockages and communication link disruptions, which poses challenges in UAV-aided systems.
To directly combat the impact of varying elevation angles, this paper proposes an adaptive phase-shift configuration-enabled RIS, which mitigates blockage effects in the A2G communication and ensures a reliable communication link.
    This letter's key contributions are as follows:

\begin{itemize}
    \item We propose a technical framework that assigns the non-central chi-square (NCCS) distribution of Rician fading channel power gain to the sum of independent shadowed Rician (SISR) distribution.
    \item We propose a stochastic A2G channel model that integrates both blockage and non-blockage conditions. The blockage condition represents high attenuation scenarios due to obstructions, which leads to A2G link disruptions.  
    \item With our framework, tractable A2G and G2A SINR results are provided in a Rician fading environment in the presence of non-IID Co-Channel Interference (CCI).
    \item Numerical results indicate that larger RIS elements, specifically 256, not only improve end-to-end (e2e) OP, but also effectively mitigate blockage impact.
\end{itemize}

{\noindent \it Notations}: Vectors are represented by boldface lowercase letters. Matrices are denoted by boldface uppercase letters. 
    The transpose, conjugate transpose, and diagonal matrices are denoted as $(\cdot)^{\sf T}$, $(\cdot)^{\sf H}$, $\operatorname{diag}(\cdot)$, respectively.
    ${ {\cal CN}(0, \sigma^2) }$ denotes the circularly symmetric complex Gaussian Random Variable~(RV) with zero mean and variance $\sigma^2$,
    ${ \mathbb{E}[\cdot] }$ denotes the expectation operator,
    ${ \mu_{k, X} = \mathbb{E}[X^k] }$ denotes the $k$th moment of~RV $X$, and ${ \mathfrak{L}_{X}(s) = \mathbb{E}[e^{-sX}] }$ denotes the Laplace transform of~$X$.

\section{System Model and Proposed Markov Process-based Phase-Shift Configuration} \label{section_system}

\begin{figure} 
	\centering
	\includegraphics[width= 0.65 \linewidth]{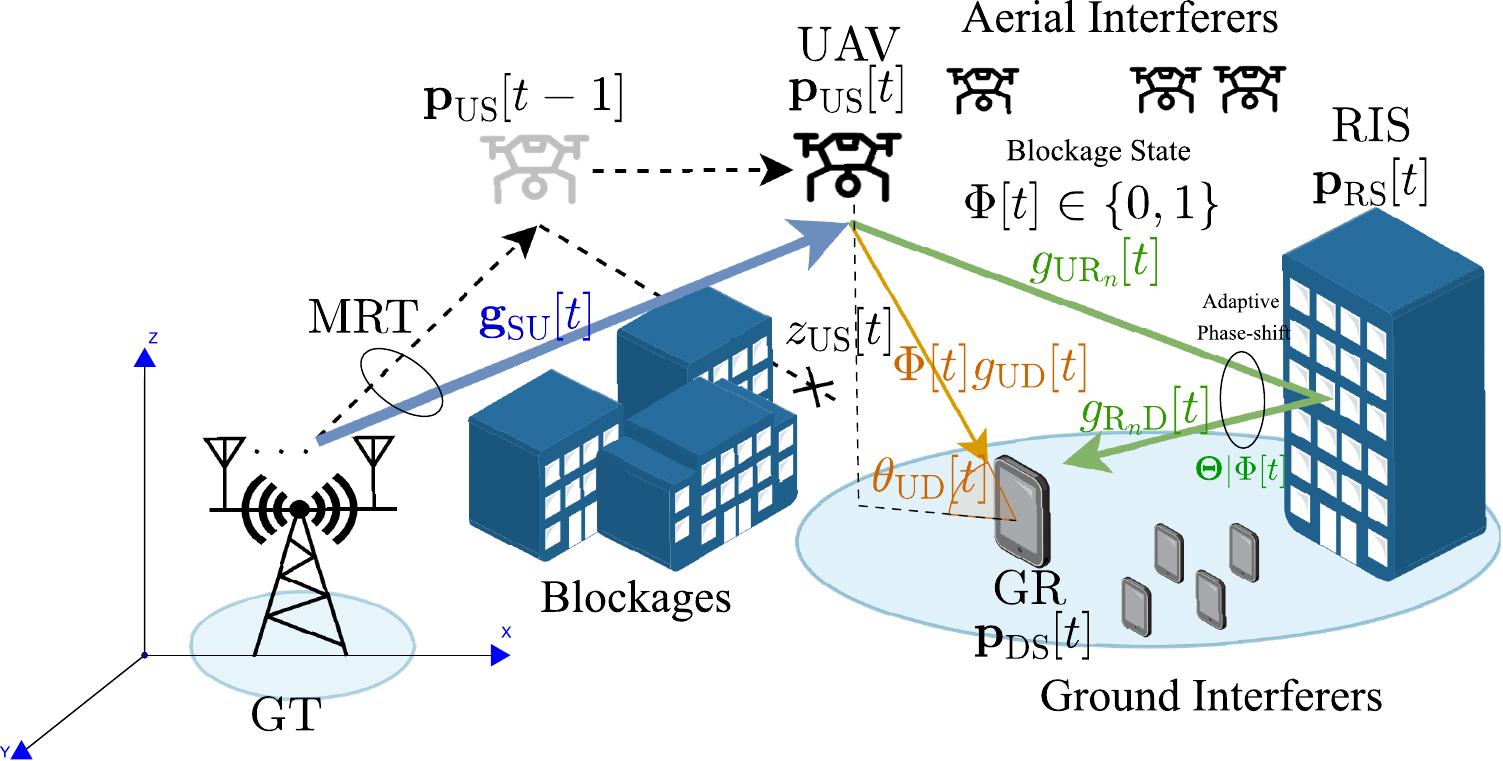}
	\caption{Illustrations of a UAV-RIS-aided wireless system in the presence of non-IID aerial and ground CCIs.}
	\label{fig1_system}
\vspace{-\baselineskip}
\end{figure}

We consider an interference-limited UAV-RIS-aided system involving an $M$-antenna ground transmitter (GT), a single-antenna UAV, a single-antenna ground receiver (GR), and an $N = N_x \times N_y$ element RIS, as depicted in Fig.~\ref{fig1_system}. 
    The system contends with $L$ aerial and $K$ ground interferers impacting G2A and A2G communications. Hereafter, we use $\src$, $\src_m$, $\uav$, $\ris$, $\ris_n$, $\des$, $\ccia_l$, and $\ccig_k$ to represent the GT, its $m$-th antenna, the UAV, the RIS, RIS's $n$-th element, the GR, the $l$-th aerial, and the $k$-th ground interferer, respectively.

We consider that the direct links $\src_m$-$\des$, ${\forall m}$, are unavailable due to multiple impenetrable obstacles, such as concrete-walled structures in the 3D-urban micro (UMi) environment.
    The relative location of node ${\nodeX}$ with respect to node ${\nodeY}$, for ${\nodeX, \nodeY \!\in\! \{ \src_m, \uav, \des, \ris_n, \ccia_l, \ccig_k \}}$, in the three-dimensional (3D) Cartesian coordinates system is denoted~as ${ {\bf p}_{\XY} \triangleq (x_{\XY}, y_{\XY}, z_{\XY}) }$, and the corresponding spherical coordinates are ${ (d_{\XY}, \theta_{\XY}, \varphi_{\XY}) }$, where $d_{\XY}$, ${ \theta_{\XY} }$, and ${ \varphi_{\XY} }$ are the radial distance, the elevation angle, and the azimuth angle, respectively. 
The channel coefficient between nodes $\nodeX$ and $\nodeY$ is denoted by ${g_{\XY} = \sqrt{\pl_\XY} h_{\XY} e^{j2\pi f_{d, \max} \cos\varpi_\XY t}}$, where $h_{\XY}$ is the complex small-scale fading, $\pl_\XY$ represents the path loss, $f_{d, \max}$ [Hz] is the maximum Doppler frequency shift, $\varpi_\XY$ [rad] is the Angle-of-Arrival (AoA) angle, and $t$ is the discrete-time instance. We consider that the UAV is moving with velocity ${\vec{v} \in \mathbb{R}^3}$. 
Herein, $\Vert \vec{v} \Vert$ [m/s] denotes the UAV's movement speed and $\frac{\vec{v}}{\Vert \vec{v} \Vert}$ specifies the unit-size movement direction. The maximum Doppler frequency shift and AoA cosine are given by ${f_{d, \max} \!=\! \frac{f_c \Vert \vec{v} \Vert}{c} }$ with $f_c$ [Hz] being the carrier frequency, and ${\cos\varpi_\XY \!=\! \frac{\vec{v}^T \vec{p}_{\XY}}{\Vert \vec{v} \Vert \Vert \vec{p}_{\XY} \Vert} }$, respectively, with $c$ [m/s] being the speed of light. For simplicity, we assume ${t \!=\! 1}$ throughout the rest of the paper.
In addition, we assume that all channels experience Rician-$K$ fading, where the small-scale fading is modeled as ${ h_{\XY} \!=\! \frac{ \sqrt{K_{\XY}} \check{h}_{\XY} + \tilde{h}_{\XY} }{ \sqrt{K_{\XY}+1} } }$ with $\check{h}_{\XY}$ being the deterministic LoS component with unit modulus and $\tilde{h}_{\XY}$ representing the scattering component.
    The $K$-factor is defined as  ${K_{\XY} \!=\! K_1 e^{K_2 \theta_{\XY}}}$, where ${ \theta_{\XY} \!=\! \sin^{-1}\frac{|z_{\XY}|}{d_{\XY}} }$ [rad] is the elevation angle of ${\nodeX}$ relative to ${\nodeY}$, and $K_1$ and $K_2$ are determined by the environment~\cite{YouTWC2019}.
The power gain of the ${\nodeX\text{-}\nodeY}$ channel, i.e., $|g_{\XY}|^2$, is an NCCS RV with the following PDF \cite{Peebles2000}:
\begin{align}
f_{|g_{\XY}|^2} (z) 
    =  e^{-\lambda_\XY z - K_{\XY} } 
    \lambda_\XY
    I_0 \big(2 {\tsqrt{K_{\XY}\lambda_\XY z}} \big),
\label{eq:pdf_gXY}
\end{align}
for ${z \!>\! 0}$, where ${ \lambda_\XY \triangleq \frac{K_{\XY} + 1}{{\pl}_{\XY}} }$ and $I_{v}(x)$ is the $v$-th order modified Bessel function of the first kind \cite{Gradshteyn2007}.
 
\subsubsection{Ground-to-Air (G2A) Communication}

Since the $\src_m$-$\des$ links are unavailable, $\src$ steers the beamforming vector $\vec{w} \in \mathbb{C}^{M}$ towards $\uav$. Hence, the received signal at $\uav$ is given by
\begin{align}
y_\uav
    = \sqrt{P_\src} 
    \vec{g}_{\src \uav}^{\sf H} \vec{w} s
    + \!\! \tsum_{l = 1}^{L}{ \tsqrt{P_{\ccia_l} } g_{\ccia_l \uav} s_{\ccia_l} } 
    + \!\! \tsum_{k = 1}^{K}{ \tsqrt{P_{\ccig_k}} g_{\ccig_k \uav} s_{\ccig_k} } + n_\uav,
\label{eq_reiSig_atUAV}
\end{align}
where ${\vec{g}_{\src\uav} \!=\! [g_{\src_1\uav}, \ldots, g_{\src_M\uav}]^{T} \in \mathbb{C}^{M \times 1}}$ is the complex channel vector from $\src$ to $\uav$ with ${g_{\src_1\uav}, \ldots, g_{\src_M\uav}}$ being statistically independent and identically distributed (IID),
    $P_\src$, $P_{\ccia_l}$, and $P_{\ccig_k}$ are the transmit powers of $\src$, the $l$-th aerial interferer, and the $k$-th ground interferer, respectively, $s$, $s_{\ccia_l}$, and $s_{\ccig_k}$ are unit-energy signals of the respective elements, and ${n_\uav}$ is the complex white Gaussian noise (AWGN) with zero mean and variance~$\sigma_\uav^2$.

\subsubsection{Air-to-Ground (A2G) Communication}

After receiving and decoding the information signal from $\src$, $\uav$ sends $\hat{s}$ to $\des$, which is reflected by $\ris$.
    Unlike most works, we consider that the $\uav$-$\des$ link, i.e., the direct A2G link, is stochastically {\it blocked}.
Specifically, the received signal strength over a blocked link is below an acceptable threshold due to propagation conditions, which is different from the NLoS state where the signal received via the NLoS paths may still be utilized for reception. 

    We consider that the blockage state of the direct A2G link follows a discrete-time Markov process ${\chain = \{ \phi_0, \phi_1, \dots, \phi_{\infty} \}}$, where $\phi_t \!\in\! \{ 0, 1 \}$, ${\forall t \ge 0}$, is the instantaneous channel state, ${\phi_t \!=\! 0}$ (blocked) for being disrupted due to blockage and ${ \phi_t \!=\! 1}$ (unblocked) for the LoS/NLoS states. The process's transition matrix is ${ [{\bf P}]_{ij} = p_{ij} }$, where $p_{ij}$ denotes the probability of transitioning from state ${i}$ to state~${j}$.
    In addition, let $\pi_0$ and $\pi_1$ be the steady-state probabilities of the blockage and non-blockage states, respectively, and ${ {\boldsymbol\pi} = [\pi_0, \pi_1]^{\sf T} }$,
the relationship between ${\boldsymbol\pi}$ and the transition matrix can be expressed as ${ {\boldsymbol\pi}^{\sf T} = {\boldsymbol\pi}^{\sf T} {\bf P} }$ \cite[Eq. (9.13)]{Peebles2000}.
    Hence, 
\begin{align}
    \pi_0 = \frac{p_{10}}{p_{01}+p_{10}}, \quad 
    \pi_1 = \frac{p_{01}}{p_{01}+p_{10}}, \quad
    \pi_0 + \pi_1 = 1.
\end{align}

In addition, the considered Markov chain $\Phi$ satisfies the following memoryless property:
\begin{align}
&\Pr(\phi_t \mid \phi_{t-1}, \phi_{t-2}, \dots) 
\nonumber\\
&\quad
= \Pr(\phi_t \mid \phi_{t-1})
=	\Bigg\{ \!\!\!
    \begin{array}{cl}
        \beta, &  \phi_{t-1} > \phi_t, \\
        \beta^{1-\phi_t} (1-\beta)^{\phi_t}, & \phi_{t-1} = \phi_t, \\ 
        1 - \beta, & \phi_{t-1} < \phi_t,
    \end{array} 
\label{eq:markov_memless_2}
\end{align}
where $\beta$ denotes the blockage probability, which will be detailed in Section \ref{subsec:blockage_prob}. The received signal at $\des$ is
\begin{align}
y_\des
    &=   \tsqrt{P_\uav} \vec{g}_{\uav\ris}^{\sf T} \vec{\Theta} \vec{g}_{\ris\des} \hat{s}
    +   \tsum\nolimits_{k = 1}^{K}{ \tsqrt{P_{\ccig_k}} g_{\ccig_k \des} s_{\ccig_k} } 
\nonumber\\
    &\quad 
    +   \tsqrt{P_\uav} \check{g}_{\uav\des} \hat{s} 
    \!+\!   \tsum\nolimits_{l = 1}^{L}{\! \tsqrt{P_{\ccia_l}} g_{\ccia_l \des} s_{\ccia_l} } 
    \!+\!   n_\des,\!\!\!
\label{eq_received_sign}
\end{align}
where ${\check{g}_{\uav\des} \!\triangleq\! \phi_0 g_{\uav\des}}$ is the direct A2G channel coefficient under the blockage effect,
    $P_\uav$ is the UAV's transmit power,
    $\vec{g}_{\uav\ris}, \vec{g}_{\ris\des} \in \mathbb{C}^{M \times 1}$ are the incident and reflected complex channel vectors from $\uav$ to~$\ris$ and from $\ris$ to $\des$, respectively, 
    ${ n_\des \sim \mathcal{CN}(0, \sigma_\des^2) }$ is the AWGN, 
    ${\bf \Theta} \triangleq \mathrm{diag}([\kappa_1 e^{j\vartheta_1}, \ldots, \kappa_N e^{j \vartheta_N}]) \in \mathbb{C}^{N \times N}$ is the phase-shift matrix of $\ris$, where ${\kappa_n \!\in\! (0,1]}$ and ${\vartheta_n \!\in\! [0, 2\pi)}$ denotes the amplitude reflection coefficient and the phase-shift of the $n$-th reflecting element of $\ris$. In addition, we assume that ${ \kappa_1 \!=\! \ldots \!=\! \kappa_N \!=\! \kappa }$~\cite{DoTCOM2021, BoulogeorgosTVT2022}.
    
\subsubsection{SINR Modeling}
From \eqref{eq_reiSig_atUAV} and \eqref{eq_received_sign}, the G2A and A2G SINRs are respectively formulated as
$\sinr_\gtoa
    \!=\!  \frac{ \frac{P_\src}{\sigma^2_\uav} \Vert \vec{g}_{\src\uav}^{\sf H} \vec{w} \Vert^2 }
            {\gamma_{\ccia\uav} + \gamma_{\ccig\uav} + 1}$ and 
$\sinr_\atog
    \!=\!  \frac{ \frac{P_\uav}{\sigma^2_\des} \left| {\bf g}_{\uav\ris}^{\sf T} \vec{\Theta} {\bf g}_{\ris\des} + \check{g}_{\uav\des} \right|^2 }
            {\gamma_{\ccia\des} + \gamma_{\ccig\des} + 1}$,
where 
    ${ \gamma_{\ccia\nodeX} \!\triangleq\! \sum_{l = 1}^{L}  \!\! \bar{\gamma}_{\ccia_l\nodeX} |h_{\ccia_l\nodeX}|^2 }$ and
    ${ \gamma_{\ccig\nodeX} \!\triangleq\! \sum_{k = 1}^{K} \!\! \bar{\gamma}_{\ccig_k\nodeX} |h_{\ccig_k\nodeX}|^2 }$
are the aggregated aerial and ground Interference-to-Noise Ratios (INRs) at ${\nodeX \!\in\! \{ \uav, \des \}}$, respectively,
    ${ \bar{\gamma}_{\ccia_l\nodeX} \!\triangleq\! \frac{P_{\ccia_l} \pl_{\ccia_l\nodeX}}{\sigma_\nodeX^2} }$, ${l \!\in\! [1, L]}$, and 
    ${ \bar{\gamma}_{\ccig_k\nodeX} \!\triangleq\! \frac{P_{\ccig_k} \pl_{\ccig_k\nodeX}}{\sigma_\nodeX^2} }$, ${k \!\in\! [1, K]}$.
The e2e SINR is determined via the weakest communication link, specified by the minimum between the G2A and A2G SINRs, and is formulated as $\sinr_{\rm e2e} \triangleq \min\{ \Gamma_{\gtoa}, \Gamma_{\atog} \}$.

    To obtain ${ \{ {\bf w}^\star, {\bf \Theta}^{\star} \} = \operatorname{arg} \max_{ {\bf w}, {\bf \Theta} }~{\Gamma_{\rm e2e}} }$, we use the fact that $\max_{ {\boldsymbol \theta} } \{ \min \{A_{\boldsymbol \theta}, B_{\boldsymbol \theta}\}\} \!=\! \min\{\max_{\boldsymbol \theta} \{A_{\boldsymbol \theta}\}, \max_{\boldsymbol \theta}\{B_{\boldsymbol \theta}\}\}\}$. 
    As the SINRs at $\uav$ and $\des$ are independent of ${\bf \Theta}$ and ${\bf w}$, respectively, the optimal beamforming vector ${\bf w}$ and phase-shift matrix ${\bf \Theta}$ that maximize the SINRs at $\uav$ and $\des$ also maximize the e2e SINR.
As a result, we use the optimal active beamforming, i.e., Maximum-Ratio Transmission (MRT) beamforming, where the $m$-th weight of the beamforming is given by
${ [\vec{w}]_m \!=\! \frac{\left| [\vec{g}_{\src\uav}]_m\right|}{\Vert \vec{g}_{\src\uav} \Vert} e^{j\angle [\vec{g}_{\src\uav}]_m} }$.
    To maximize the A2G SINR, we use the fact that $\big| {\bf g}_{\uav\ris}^{\sf T} \vec{\Theta} {\bf g}_{\ris\des} + \check{g}_{\uav\des} \big| \le \big| \sum_{n=1}^{N}{|g_{\uav\ris_n} g_{\ris_n\des}|} + |\check{g}_{\uav\des}| \big|$ where the equality occurs when $\vartheta_n + \angle g_{\uav\ris_n} + \angle g_{\ris_n\des} = \angle \check{g}_{\uav\des}$ for all $n \in [1, N]$.
It is noted that the phase-shift configuration depends on the availability of the direct A2G link. 
Hence, we obtain the following adaptive phase-shift configuration:
\begin{align}
&\vartheta^{\ast}_n 
   = \phi_0 \big[
        \varphi_{\uav\des} 
            + 2 \pi f_{d, \max} {\vec{v}^T \vec{p}_{\uav\ris_n}}/({\Vert \vec{v} \Vert \Vert  \vec{p}_{\uav\des} \Vert})
        \big]
    \nonumber\\
    &\hspace{20pt}
        - \varphi_{\uav\ris_n} 
        \!-\! \varphi_{\ris_n\des} 
        \!-\! 2 \pi f_{d, \max} {\vec{v}^T \vec{p}_{\uav\ris_n}}/({\Vert \vec{v} \Vert \Vert \vec{p}_{\uav\ris_n} \Vert}).
\end{align}
By adaptively adjusting $\vartheta^\ast_n$ based on the instantaneous state of the direct A2G link, the proposed two-state Markov process-based adaptive phase shift can maintain reliable communication, even in the presence of blockages.
Subsequently, the optimal e2e SINR is obtained as ${\sinr_{\rm e2e}^\star \!=\! \min\{ \sinr_\gtoa^\star, \sinr_\atog^\star \}}$ where
\vspace{-8pt}
\begin{align} 
\sinr_\gtoa^\star
    \!=\!  \frac{ \bar{\gamma}_{\gtoa} \gamma_{\gtoa} }
        { \gamma_{\ccia \uav} \!+\! \gamma_{\ccig \uav} \!+\! 1},
\sinr_\atog^\star
    \!=\!  \frac{ \big| \sqrt{ \bar{\gamma}_{\cas} \gamma_{\cas} } +  \sqrt{\bar{\gamma}_{\dir} \gamma_{\dir}} \big|^2 }
        { \gamma_{\ccia \des} \!+\! \gamma_{\ccig \des} \!+\! 1},
\end{align}
where  
    ${ \bar{\gamma}_{\gtoa} \!\triangleq\! \frac{P_{\src} \pl_{\src\uav}}{\sigma^2_\uav} }$,
    ${ \gamma_{\gtoa} \!\triangleq\! \sum_{m=1}^{M}{\!\! (h_{\src_m\uav})^2} }$,
    ${ \bar{\gamma}_{\cas} \!\triangleq\! \frac{P_\uav \pl_{\uav\ris} \pl_{\ris\des} |\kappa|^2 \!\!}{\sigma^2_\des} }$, 
    ${ \bar{\gamma}_{\dir} \!\triangleq\! \frac{P_\uav \pl_{\uav\des}}{\sigma^2_\des} }$, 
    ${ {\gamma}_{\cas} \!\triangleq\! (\sum_{n=1}^{N}{\!\! h_{\uav\ris_n} h_{\ris_n\des} })^2 }$, 
    and~${ \gamma_{\dir} \!\triangleq\! \phi_0 |h_{\uav\des}|^2 }$.
    
\section{Aerial Chanel Characterization and End-to-End Performance Analysis}

Characterizing the stochastic behavior of Rician fading channels in UAV-RIS-aided communication presents mathematical challenges due to the use of Bessel functions and the presence of non-IID interference. Additionally, modeling blockage events with a two-state Markov chain further increases the complexity of the performance analysis.

\subsection{Sum of independent shadowed Rician (SISR) distribution}

To assist in the performance analysis, we propose alternative expressions of \eqref{eq:pdf_gXY} with the help of the following the Lemma.
\begin{lemma}
\label{lem:pdfcdf_X2MG}
Let ${ Y \!=\! \sum_{i=1}^{I}{ X_i } }$ be the SISR RV, where each $X_i$ is a shadowed-Rican RV whose parameters are $\kappa$, $m$, and $\omega$. The PDF and CDF of $Y$ are respectively given~by:
\begin{align}
f_{Y} (I, \psi; x) 
    &=  \!\!\!
    \tsum_{k=0}^{I \psi - I}
    \frac{ \chi_{I, k} }{ (I \psi-k-1)! } 
    \frac{x^{I \psi-k-1}}{\alpha^{I \psi-k}} e^{-\frac{x}{\alpha}},{x > 0}, \!
\label{eq_pdf_hAB_mag_RK2MG} \\
F_{Y} (I, \psi; x) 
    &=  1 - \!\!\! 
        \tsum_{k=0}^{ I \psi - I}
        \chi_{I, k} \frac{\gamma(I \psi - k, x/\alpha)}{(I \psi-k-1)!} ,{x > 0},
\label{eq_cdf_hAB_mag_RK2MG_a} \\
    &=  1 - \!\!\! 
    \tsum_{k=0}^{ I \psi - I}
    \tsum_{n=0}^{I \psi - k - 1}
    \frac{ \chi_{I, k} }{ n! } 
    \frac{x^n}{\alpha^n} e^{-\frac{x}{\alpha}},{x > 0},
\label{eq_cdf_hAB_mag_RK2MG}
\end{align}
where  $\gamma(a, x)$ is the lower incomplete Gamma function~\cite{Gradshteyn2007},
    ${ \alpha \!\triangleq\! \frac{\xi + 1}{\kappa + 1} }$,
    ${ \chi_{I, k} \!\triangleq\! \binom{I \psi-I}{k} \frac{1}{ (\xi + 1)^k } ( \frac{\xi}{\xi + 1})^{I \psi - I - k} }$,~${ \xi \!\triangleq\! \frac{\kappa\omega}{\psi} }$,~${ \kappa \!\triangleq\! \frac{1\!-\!\sigma^2}{\sigma^2} }$,
    ${ \omega \!\triangleq\! \frac{\Omega (\kappa+1)}{\kappa} }$,~
    ${ \Omega \!\triangleq\! \frac{\sqrt{\psi}}{\sqrt{\psi-1}} \frac{\sqrt{\mu_1^2(I+1)-I \mu_2}}{I} }$,~and~${ \sigma^2 \!\triangleq\! \frac{\mu_2 - I\Omega}{I} }$. 
\end{lemma}
\begin{IEEEproof}
   Provided in the Appendix \ref{apd_a}.
\end{IEEEproof}

The above Lemma allows us to represent \eqref{eq:pdf_gXY} via the SISR distribution with ${ \mu_1 \!=\! \mathbb{E}\{ |g_\XY|^2 \} }$ and ${ \mu_2 \!=\! \mathbb{E}\{ |g_\XY|^4 \} }$. 
    The parameter $I$ should be proportional to the degrees of freedom of $|g_{\XY}|^2$, e.g., ${I \!=\! 1}$ for 2 degrees of freedom, and should satisfy ${ I \!\le\! \lfloor \frac{\mu_1^2}{\mu_2-\mu_1^2} \rfloor }$.
As the LoS component is dominant, the complexity-accuracy trade-off coefficient $\psi$ should be relatively large  
to achieve high accuracy. 
    It is worth noting that the statistical nature of the channel remains Rician, not shadowed-Rician, and the above Lemma is proposed to be later utilized in Sections \ref{sec:g2a_sinr} and~\ref{sec:a2g_sinr}.

\subsection{Distribution of the G2A SINR}
\label{sec:g2a_sinr}

In this subsection, we derive the closed-form expression of the G2A SINR's CDF, which is presented in Theorem \ref{theo_G2ACDF}.

\begin{Theorem}
\label{theo_G2ACDF}
The CDF of the G2A SINR is
\begin{align}
F_{\sinr_\gtoa^\star}(x) =  
    1 - 
    \tsum_{k=0}^{M \psi_{\src\uav} - M} 
    \!\!
	\chi_{M, k} 
    \!\!\!\!\! 
    \tsum_{n=0}^{M \psi_{\src\uav} - k - 1}
    \frac{\Delta^{(n)}_{\uav}(\frac{x}{\alpha_{\gtoa} \bar{\gamma}_{\gtoa}})}{n!},
\end{align}
for $x > 0$, where $\Delta^{(n)}_{\uav}(s)$ is given by \eqref{eq:delta_Xn}.
\end{Theorem}

\begin{IEEEproof}
First, the CDF of $\sinr_\gtoa^\star$ can be derived~as
${F_{\sinr_\gtoa^\star}(x)
    \!=\! \mathbb{E}\big\{
        F_{\gamma_{\gtoa}}
        ({x (\gamma_{\ccia \uav} \!+\! \gamma_{\ccig \uav} \!+\! 1)}/{\bar{\gamma}_{\gtoa}})
    \big\} }$.
Using the proposed Lemma, we rewrite the CDF of $\gamma_{\gtoa}$ with~${I \!\triangleq\! M}$ and a pre-defined trade-off parameter ${\psi \triangleq \psi_{\src\uav} }$. Here, the $k$th moment of $|h_{\src_m\uav}|^2$ is required, where
${\mu_{k, |h_{\XY}|^2} \!\triangleq\! \frac{\Gamma(1 + k) L_{k}(-K_\XY)}{(1 + K_\XY)^k }}$ and $L_k(x)$ is the Laguerre polynomial of order $k$. 
    Moreover, the $k$th moment of $\gamma_{\gtoa}$ is determined~as
\begin{align}
\mu_{k, \gamma_{\gtoa}} 
=   \tsum_{r=1}^{k}
    \frac{k! M!}{(M \!-\! r)!}
    \mathop{\widetilde{\textstyle\sum}}\limits_{r, k}
    \bigg(
        \tprod_{j=1}^{r} { \frac{1}{p_j!} }
    \bigg)
    \tprod_{i=1}^{\varrho({\bf \Sigma})}{ \frac{ (\mu_{p_{\langle i \rangle}})^{\nu_i} }{ \nu_i! } },
\label{eq:nth_moment_gSU}
\end{align}
where $\mu_{i} \triangleq \mu_{i, |h_{\XY}|^2}$,
    ${ \mathop{\widetilde{\Sigma}}_{r, k} }$ is the shorthand notation for the summation over $r$ integers that satisfy ${ p_1 \!+\! \cdots \!+\! p_r \!=\! k }$ and ${ p_1, \dots, p_r \ge 1 }$,
    ${ {\bf \Sigma} = \operatorname{diag}([p_1, \dots, p_r]) }$, 
    $\varrho({\bf \Sigma})$ is the number of distinct diagonal elements in $\bf \Sigma$, 
    ${ p_{\langle 1 \rangle} > \cdots > p_{\langle \varrho({\bf \Sigma}) \rangle} }$ are the distinct diagonal elements of $\bf \Sigma$ in decreasing order, and
    $\nu_i$ is the multiplicity of $p_{\langle i \rangle}$.
We can further derive $F_{\sinr_\gtoa^\star}(x)$~as
\begin{align}
&F_{\sinr_\gtoa^\star}(x)
    =   1 - 
    \tsum_{k=0}^{ M \psi_{\src\uav} -M } 
    \tsum_{n=0}^{ M \psi_{\src\uav} -k-1 } 
    \chi_{M, k} 
    (-s_0)^n
\nonumber\\
    &\qquad\times
    \frac{1}{n!}
    \bigg[
    \frac{{\rm d}^n}{{\rm d}s^{n}}
        e^{-s}
        \tprod_{l=1}^{L}{ {\mathfrak{L}}_{\gamma_{\ccia_l \uav}} (s) }
        \tprod_{k=1}^{K}{ {\mathfrak{L}}_{\gamma_{\ccig_k \uav}} (s) }
    \bigg]
    \bigg|_{s\to s_0},
    \label{eq_proofTheo1_a} 
\end{align}
where ${ s_0 \triangleq \frac{x}{\alpha \bar{\gamma}_{\gtoa}} }$. 
    Using \cite[Table I]{LopezTC2017} for expanding $ {\mathfrak{L}}_{\gamma_{\ccia_l \uav}} (s) $ and $ {\mathfrak{L}}_{\gamma_{\ccig_k \uav}} (s) $, we can rewrite the above $n$-th order derivatives as follows
\begin{align}
\Delta^{(n)}_{\uav}(s) 
    &=  \frac{{\rm d}^n}{{\rm d}s^{n}}
        e^{-s}       
        \tprod_{l=1}^{L} 
            \frac{e^{ - K_{\ccia_l\uav} + \frac{K_{\ccia_l\uav}}{1+\bar{\gamma}_{\ccia_l\uav} s} }}{1 + \bar{\gamma}_{\ccia_l\uav} s}
        \tprod_{k=1}^{K} 
            \frac{e^{ - K_{\ccig_k\uav} + \frac{K_{\ccig_k\uav}}{1+\bar{\gamma}_{\ccig_k\uav} s} }}{1 + \bar{\gamma}_{\ccig_k\uav} s}
\nonumber\\
    &=  
    \Delta^{(0)}_{\uav}(s)
    \bigg\{
        1 + \! \tsum_{r = 1}^{n}
        \mathop{\widetilde{\textstyle\sum}}\limits_{r,n}
            \frac{ n!\sprod_{i=1}^{ \varrho({\bf \Sigma}) }
            {S_{p_{\langle i \rangle}}^{\nu_i}(s)}/{\nu_i!} }{ \sprod_{i=1}^{r}{p_i!} }
    \bigg\},\!\!\!
\label{eq:delta_Xn}
\end{align}
where ${S_i(s) \!=\! \sum_{l=1}^{L}{ \delta^{(i)}_{\ccia_l\uav}(s) }
        \!+\! \sum_{k=1}^{K}{ \delta^{(i)}_{\ccig_k\uav}(s) } - C_i}$
with ${C_1 \!=\! 1}$ and ${C_{j\ge 2} \!=\! 0}$, and ${\delta^{(n)}(s) \!=\! (n \!-\! 1)! \frac{(-\bar{\gamma})^n}{(1+\bar{\gamma} s)^n} (1 \!+\! \frac{n K}{1+\bar{\gamma} s})}$. The detailed derivation of \eqref{eq:delta_Xn} is omitted here due to space constraint\footnote{The full proof of Theorem 1 is available at https://github.com/thanhluannguyen/UAV-RIS-blockage.}.
This completes the proof of Theorem~\ref{theo_G2ACDF}.
\end{IEEEproof}

\subsection{Distribution of A2G SINR}
\label{sec:a2g_sinr}

In this subsection, we investigate the distribution of the A2G SINR and derive its CDF in the following Theorem.
    
\begin{Theorem}
\label{theo:CDF_A2G}
The CDF of the A2G SINR is given by
\begin{align}
\label{eq:cdf_a2g_sinr}
F_{\sinr_\atog^\star}(x)
    =   1 - \!\!\! 
    \tsum_{k=0}^{I \psi - I} 
    \!\!
	\chi_{I, k} 
    \!\!\!\!\! \tsum_{n=0}^{I \psi - k - 1}
    \frac{\Delta^{(n)}_{\des}(\frac{x}{\alpha_{\cas} \bar{\gamma}_{\cas}})}{n!}, x> 0.
\end{align}
\end{Theorem}
\begin{IEEEproof}
Let ${ \gamma_{\atog} \!\triangleq\!  \big| \sqrt{\gamma_{\cas} } +  \sqrt{\gamma_{\dir} \bar{\gamma}_{\dir}/\bar{\gamma}_{\cas}} \big|^2 }$ be the normalized A2G SNR.
Next, we adopt the proposed Lemma to approximate the distribution of~$\gamma_{\atog}$, which requires the $k$th moments~of~${\gamma_{\dir} \bar{\gamma}_{\dir}/\bar{\gamma}_{\cas}}$ and of $\gamma_{\cas}$.
    Under blockage, the PDF of $|\check{g}_{\uav\des}|^2$, defined as ${f_{|\check{g}_{\uav\des}|^2}(x) \!=\! \mathbb{E}[f_{\phi_0|g_{\uav\des}|^2}(x)] }$, is
\begin{align}
\label{eq:pdf_checkg}
f_{|\check{g}_{\uav\des}|^2}(x) 
    &=  \frac{\pi_1}{1 - \beta}
    e^{-\lambda_{\uav\des} x - K_{\uav\des} } 
    \lambda_{\uav\des} 
    I_0 \big(2 {\tsqrt{K_{\uav\des} \lambda_{\uav\des} x}} \big),
\end{align}
for ${x \!>\! \tau}$. We first derive the $k$th moment of ${\gamma \triangleq \frac{\gamma_{\dir} \bar{\gamma}_{\dir}}{\bar{\gamma}_{\cas}}}$ as
\begin{align}
{\mu_{k, \gamma} 
    \!=\! \mathbb{E}[\gamma^k] 
    \!=\! \scalebox{1.5}{$\int$}_{\tau}^{\infty}{ f_{|\check{g}_{\uav\des}|^2}(x) 
    x^k /{(\bar{\gamma}_{\cas} \sigma_{\des}^2})^k {\rm d} x }}.    
\end{align}
Then, utilizing \eqref{eq:pdf_checkg} and invoking the Nuttall-$Q$ function $Q_{m,n}(a,b)$, we obtain
\begin{align}
{\mu_{k, \gamma}
\!=\!
\frac{ \pi_1 }{ 1 - \beta }
\frac{ Q_{2k+1,0}(\sqrt{2 K_{\uav\des}},
\sqrt{ 2 \lambda_{\uav\des}\tau }) }{ (2 \lambda_{\uav\des} \sigma_\des^2 \bar{\gamma}_{\cas})^k }}.
\end{align}
Next, we use \eqref{eq:nth_moment_gSU} to calculate the $k$th moment of $\gamma_{\cas}$ as
\begin{align}
{\mu_{k, \gamma_{\cas}} \!=\! \tsum_{r=1}^{2k}
\frac{(2k)! N!}{(N - r)!}
\mathop{\widetilde{\tsum}}_{r, 2k}
\Big( \tprod_{j=1}^{r} { \frac{1}{p_j!} } \Big)
\tprod_{i=1}^{\varrho({\bf \Sigma})}{ \frac{ (\mu_{p_{\langle i \rangle}})^{\nu_i} }{ \nu_i! } } }, 
\end{align}
where, in this case, $\mu_{i} \triangleq \mu_{\frac{i}{2}, |h_{\uav\ris_n}|^2} \mu_{\frac{i}{2}, |h_{\ris_n\des}|^2}$.

Afterwards, we rewrite the CDF of $\gamma_{\atog}$ using the Lemma \ref{lem:pdfcdf_X2MG}, which requires its $k$th moment to be
\begin{align}
{\mu_{k, \gamma_{\atog}} \!=\! \tsum_{i=0}^{2k} {\textstyle \binom{2k}{i}} (\mu_{k-\frac{i}{2}, \gamma_\cas}) { (\mu_{\frac{i}{2}, \gamma}) }}.
\end{align}

After a series of mathematical manipulations, we obtain~\eqref{eq:cdf_a2g_sinr}. This completes the Proof of Theorem~\ref{theo:CDF_A2G}.
\end{IEEEproof}

\subsection{Asymptotic analysis}
\label{subsec:asymptotic}

We have ${ F_{Y}(I, \psi; x) \!\to\! \sum_{k=0}^{ I \psi - I} \frac{\chi_{I, k}}{(I \psi-k)!} (\frac{x}{\alpha})^{I \psi-k} }$ as ${x \!\to\! 0}$ by using \cite[Eq. (8.354.1)]{Gradshteyn2007}. 
In high transmit power regimes, where ${P_\src \!=\! P_\uav \!=\! P \!\to\! \infty}$, $F_{\Gamma_{\gtoa}^{\star}}(x)$ can be further simplified~as
\begin{align}
F_{\Gamma_{\gtoa}^{\star}}(x)
    \to
    \tsum_{k=0}^{ M \psi_{\src\uav} - M}
    \frac{\mu_{M \psi_{\src\uav} - k} \chi_{M, k}}{(M \psi_{\src\uav}-k)!}
    \Big( 
        \frac{x}{\alpha_{\gtoa} \bar{\gamma}_{\gtoa}}
    \Big)^{M \psi_{\src\uav} - k} \!\!\!\!\!\!\! \!\!\!\!\!\!\!,
\end{align}
where, for this equation, ${\mu_k \!=\! \mathbb{E}[(\gamma_{\ccia\uav} \!+\! \gamma_{\ccig\uav} \!+\! 1)^k]}$ and is derived~as ${\mu_k \!=\! (-1)^{k} [ \frac{{\rm d}^k}{{\rm d} s^k} \mathfrak{L}_{\gamma_{\ccia\uav} \!+\! \gamma_{\ccig\uav} \!+\! 1}(s) ] \vert_{s\to 0} \!=\! (-1)^{k} \Delta_{\uav}^{(k)}(0)}$. 

Hence, applying the above steps, $F_{\Gamma_{\atog}^{\star}}(x)$ when ${P \!\to\! \infty}$ can be simplified~as
\begin{align}
F_{\Gamma_{\atog}^{\star}}(x) \to
\tsum_{k=0}^{ I \psi - I}
    \frac{\mu_{I \psi-I} \chi_{I, k}}{(I \psi-I)!}
    \Big( 
        \frac{x}{\alpha_{\atog} \bar{\gamma}_{\atog}}
    \Big)^{I \psi - k},
\label{eq:cdf_atog_asy}
\end{align}
where, for \eqref{eq:cdf_atog_asy}, ${\mu_k = \mathbb{E}[(\gamma_{\ccia\des} + \gamma_{\ccig\des} + 1)^k] = (-1)^{k} \Delta_{\des}^{(k)}(0)}$.

\subsection{Blockage probability of direct A2G channel}
\label{subsec:blockage_prob}

The Markov stochastic process serves as a dynamic representation of the direct A2G link's availability. 
    In scenarios where the direct A2G link encounters a blockage (${ \phi_0 = 0 }$), the value of the direct A2G channel power gain $|g_{\uav\des}|^2$ falls below a predefined threshold $\tau$, rendering it unavailable for communication purposes. 
In this case, the e2e communication is redirected through a cascaded link. 
    From \eqref{eq:markov_memless_2}, the blockage probability can be obtained~as
\begin{align}
\beta 
    &=\Pr(|g_{\uav\des}|^2 < \tau)
    \\
    &=  1 - Q\big(
        \sqrt{2 K_{\uav\des}},
        \sqrt{ 2 \lambda_{\uav\des} \tau }
    \big),
\end{align}
where $Q(a, b)$ is the first order Marcum Q-function.

The impact of the blockage probability on the system performance is presented in the following section.

\section{Numerical Results} \label{sec:numerical_result}

In this section, we validate the accuracy of our derived analytical expressions based on numerous numerical simulations. Here, we consider a noise spectral density of $-174$ dBm/Hz and system bandwidth of $10$ MHz \cite{DoTCOM2021}. We consider that $\src$ is located at the origin of a normalized 3D Cartesian coordinate system with the coordinates of $\uav$, $\des$, and $\ris$ being $(0.5, 0.5, 1.0)$, $(0.5, 0.5, 0)$, and $(1.0, 1.0, 0)$, respectively. We consider the G2A/A2G path loss model as ${ \ell_{\XY} \!=\! - G_{t} \!-\! G_{r} \!+\! 22.7 \!+\! 26\log(f_c) \!-\! 36.7\log(d_{\XY}) }$ dB, where ${ G_t \!=\! G_r \!=\! 0 }$ dBi as transmit and receive antenna gains, and ${ f_c \!=\! 3 }$ GHz as the carrier frequency \cite[Table B.1.2.1-1]{NOMA3GPP}, and ${ \ell_{\ris\des} \!=\! \!-\! G_{t} \!-\! G_{r} \!-\! 37.3  + 26\log(f_c) - 36.7\log(d_{\ris\des})~\text{dB}}$. With Rician fading, we have ${K_1 \!=\! 0}$ dB and~${K_2 \!=\! \frac{2}{\pi}\log(\frac{K_\pi}{K_0})}$, where ${K_\pi \!=\! 5}$ dB~\cite{YouTWC2019}. Additionally, we set 
    ${ {\bf p}_{\ccig_1\src} \!=\! (0.4, 0.4, 0) }$, 
    ${ {\bf p}_{\ccig_2\src} \!=\! (0.4, 0.8, 0) }$,
    ${ {\bf p}_{\ccia_1\src} \!=\! (0.2, 0.2, 0.6) }$, 
    ${ {\bf p}_{\ccia_2\src} \!=\! (0.4, 0.8, 0.4) }$, and ${P_{\ccia_l} \!=\! P_{\ccig_k} \!=\! 0}$ dBm for all ${l \!\in\! [1,L]}$ and ${k \!\in\! [1, K]}$.

\begin{figure*}[htbp]
  \centering
  \label{fig:figures}
  \begin{minipage}{0.4\linewidth}
    \centering
    \subfloat[]{
    \includegraphics[width=0.45\linewidth]{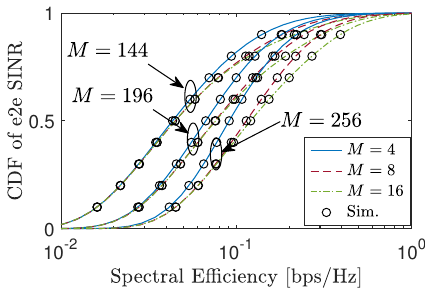}
        \label{fig2}}
    \subfloat[]{
    \includegraphics[width=0.45\linewidth]{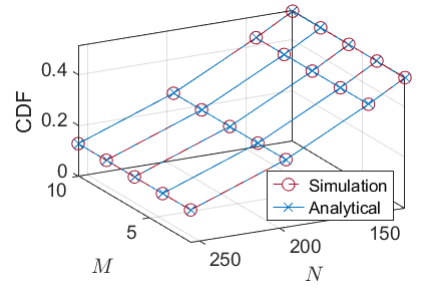}
        \label{fig3}}
    \caption{CDF of the e2e SINR as a function of a) the target SE [bps/Hz] and the number of b) transmit antenna ($M$) and RIS element ($N$). \\ \\}
  \end{minipage}
  \hfill
  \begin{minipage}{0.18\linewidth}
    \centering
    \includegraphics[width=\linewidth]{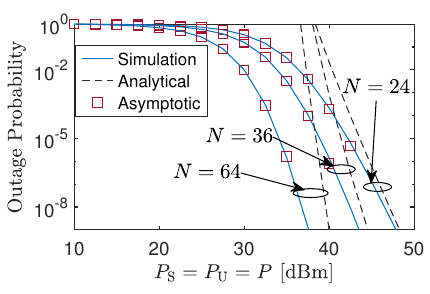}
    \caption{e2e OP versus the transmit powers versus the number of RIS elements. \\ \\ \label{fig4}}
  \end{minipage}
  \hfill
  \begin{minipage}{0.18\linewidth}
    \centering
    \includegraphics[width=\linewidth]{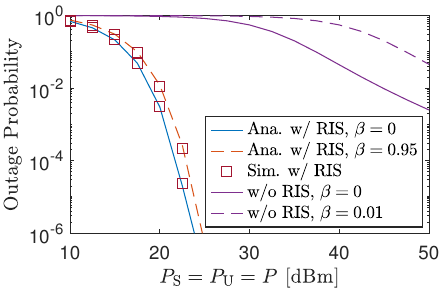}
    \caption{e2e OP with (w/) and without (w/o) the help of RIS under varying transmission powers.  \vspace{5pt} \label{fig5}}
  \end{minipage}
  \hfill
  \begin{minipage}{0.18\linewidth}
    \centering
    \includegraphics[width=\linewidth]{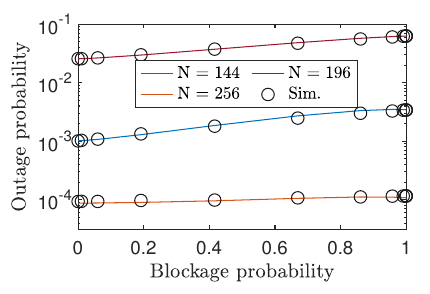}
    \caption{e2e OP versus the blockage probability ($\beta$). \vspace{25pt} \label{fig6}}
  \end{minipage}
\vspace{-30pt}
\end{figure*}

Fig. \ref{fig2} and Fig. \ref{fig3} demonstrate the close match between analytical and simulated results across different numbers of transmit antennas ($M$) and RIS elements ($N$). It is noted that we set ${P_\src \!=\! P_\uav \!=\! 0}$ dBm for normalization purposes. 
Moreover, the CDF of the e2e SINR is depicted in Fig. \ref{fig3}, as a function of $M$ and $N$ at a spectral efficiency ($R_{\tt SE}$) of 0.5 bps/Hz, ${P_\src \!=\! P_\uav \!=\! 23}$ dBm, and $\beta \!=\! 0$. Notably, increasing the number of transmit antennas or RIS elements decreases the SINR's CDF, suggesting the system is better suited for harsher environment conditions and stricter QoS requirements.
    In both figures, the analytical and simulation results match well, which validates the accuracy of our analysis. 

Fig. \ref{fig4} presents the relationship between e2e OP and the transmission power. It is noted that the e2e OP is formulated~as
${{\rm OP}_{\rm e2e} \!=\! 1 - (1-F_{\Gamma^\star_{\gtoa}}(\tau_{\rm e2e}))(1-F_{\Gamma^\star_{\atog}}(\tau_{\rm e2e}))}$.
With an increase in transmit power, the e2e OP generally decreases. Comparing the decay rate among the three configurations, we observe that $N = 64$ elements shows the fastest decay, whereas $N = 24$ elements shows the lowest. 

Fig. \ref{fig5} demonstrates the RIS's notable effect on the e2e OP across various transmission powers, where ${R_{\tt SE} \!=\! 0.5}$ bps/Hz, $N = 6$, and $M = 144$. 
    The use of RIS evidently reduces OP in both unblocked and highly blocked (blockage probability of 0.95) scenarios. Without RIS (w/o RIS), a minor blockage probability of 0.01 considerably escalates OP, underscoring RIS's critical role in system resilience and performance amidst non-negligible blockages.

In Fig. \ref{fig6}, where ${M \!=\! 4}$ and ${R_{\tt SE} \!=\! 1}$ bps/Hz, we observe that as $\beta$ increases from 0 to 1, the OP generally increases, resulting in lower e2e communication reliability. However, such a reduction in OP is alleviated by deploying more RIS elements. Noticeably, increasing $N$ to 196 yields OP of only 0.01\%.   
Furthermore, the impact of blockage is mitigated at high values of $N$, especially in the range of 144 to 196 elements.
When the number of RIS's elements is relatively large, the OP stabilizes.
In this regard, the RIS demonstrates its robustness against blockages as it guarantees stable and reliable performance with large numbers of elements.

\section{Conclusion}

In this letter, we presented stochastic A2G channel models for UAV-RIS-aided systems. Specifically, using a discrete-time Markov process, the A2G communication with ground UEs under varying conditions is modeled, including blocked and non-blockage scenarios. 
     We provided tractable closed-form expressions for the CDF of the G2A and A2G SINRs in Rician fading environments in the presence of non-IID CCI by adopting the proposed technical framework that matched the distribution of the channel power gain to the sum of independent shadowed Rician distribution. 
The results demonstrated that increasing the number of RIS elements significantly improves the e2e OP and mitigates the impact of blockage in the direct A2G~link.

\appendices
\section{Proof of the Lemma} \label{apd_a}
Using the physical model of the shadowed Rician distribution, a signal power $X_i$ can be expressed in terms of the in-phase and quadrature-phase components of the fading signals as ${X_i = (P_i + \xi_i p_i)^2 + (Q_i + \xi_i q_i)^2}$, where 
    $P_i$~and~$Q_i$ are IID Gaussian RVs with zero mean and variance $\frac{\sigma^2}{2}$; 
    $p_i$ and $q_i$ are real numbers satisfying ${p_i^2 \!+\! q_i^2 \!=\! 1}$; and $\xi_i$ is a ${\text{Nakagami-}m}$ RV with shaping parameter~$\psi$ and scale~$\Omega$.
The first and the second moments of $X_i$ are given by ${ \mathbb{E}[X_i] = \Omega + \sigma^2 }$ and ${ \mathbb{E}[X_i^2] = 2 \sigma^4 + 4 \sigma^2 \Omega + \frac{\mu+1}{\mu} \Omega^2 }$, respectively.
For chosen $I$ and the trade-off parameter $\psi$, by solving the following system of equations i) ${(\Omega \!+\! \sigma^2) Q \!=\! \mu_1}$ and ii) ${(2 \sigma^4 \!+\! 4 \sigma^2 \Omega \!+\! \frac{\psi \!+\! 1}{\psi} \Omega^2) I + I (I \!-\! 1) (\Omega \!+\! \sigma^2)^2 \!=\! \mu_2}$, where $\mu_k \triangleq \mathbb{E}[Y^k]$, we obtain $\Omega$ and $\sigma^2$ as follows
\begin{align}
\Omega &= \tsqrt{\mu_1^2(I+1)-I\mu_2} \sqrt{\psi/(\psi-1)}/I, \\
\sigma^2 &= (\mu_1 - \tsqrt{\mu_1^2(I+1)-I\mu_2} \sqrt{\psi/(\psi-1)})/I.
\end{align}

The parameters $\kappa$ and $\omega$ of shadowed Rician are obtained as $\kappa =  1/\sigma^2 - 1$ and $\omega = \Omega (\kappa+1)/\kappa$, respectively.
The rest of the analysis, including the CDF and PDF of $Y$, is presented in~\cite{AlfanoTWC2007}. This completes the proof of the Lemma.

\bibliographystyle{IEEEtran}
\bibliography{main}

\end{document}